\documentclass{amsproc}
\usepackage{epsfig}

\newtheorem{theorem}{Theorem}[section]
\newtheorem{lemma}[theorem]{Lemma}

\theoremstyle{definition}
\newtheorem{definition}[theorem]{Definition}
\newtheorem{example}[theorem]{Example}

\theoremstyle{remark}

\newtheorem{problem}[theorem]{{\bf Problem}}

\numberwithin{equation}{section}

\begin{document}

\title{An Introduction to Reconstructing Ancestral Genomes}

\author{Lior Pachter}
\address{Departments of Mathematics and Computer Science, UC Berkeley.}
\thanks{The author was supported in part by NSF Grant CCF-0347992.}

\date{\today}


\begin{abstract}

Recent advances in high-throughput genomics technologies have resulted
in the sequencing of large numbers of (near) complete genomes. These
genome sequences are being mined for important functional elements,
such as genes. They are also being compared and contrasted
in order to identify other functional sequences, such as those involved
in the regulation of genes. In cases where DNA sequences from different organisms can be
determined to have originated from a common ancestor, it is natural to
try to infer the ancestral sequences. The reconstruction of 
ancestral genomes can lead to insights about genome evolution, and the
origins and diversity of function.  There are a number of
interesting foundational questions associated with reconstructing
ancestral genomes: 
Which statistical models for
evolution should be used for making inferences about ancestral
sequences? How should extant genomes be compared in order to facilitate
ancestral reconstruction? Which portions of ancestral genomes can be
reconstructed reliably, and what are the limits of ancestral
reconstruction? We discuss recent progress on some of these questions,
offer some of our own opinions, and highlight interesting mathematics,
statistics, and computer science problems.
\end{abstract}

\maketitle

\section{What is comparative genomics?}

These notes summarize a lecture at a special session of the American
Mathematical Society on
mathematical biology, during which we discussed the central problem of
comparative genomics, namely how to reconstruct the ancestral 
genomes that evolved into the present-day extant genomes.
This is fundamentally a statistics problem, because with a few exceptions, 
it is not possible to sequence the genomes of ancestral species, and
one can only 
infer ancestral genomes from the multitude of genomes that can be
sampled at the present time. The problem is a grand scientific challenge
that has only begun to be tackled in recent years, now that whole
genomes are being sequenced for the first time.
Our aim is to introduce the reader to the statistical
(and related mathematical) elements of the methods of comparative
genomics, while providing a glimpse of the exciting results that are
emerging from first generation attempts to reconstruct ancestral genomes.
Due to the complex interdisciplinary scope of the subject, we have been forced
to omit
a lot of detail and many interesting
topics, but we hope that the curious mathematical reader may find some
threads worthy of further exploration. 

We begin with a concrete example of one sequence from a single genome: the
16S ribosomal RNA (rRNA) gene from Salmonella typhimurium LT2. This sequence
can be downloaded from the NCBI website  at {\tt
  http://www.ncbi.nlm.nih.gov/} by searching for the accession number
``AE008857''. The sequence is 

{\small
\begin{verbatim}
   1 aattgaagag tttgatcatg gctcagattg aacgctggcg gcaggcctaa cacatgcaag
  61 tcgaacggta acaggaagca gcttgctgct tcgctgacga gtggcggacg ggtgagtaat
 121 gtctgggaaa ctgcctgatg gagggggata actactggaa acggtggcta ataccgcata
 181 acgtcgcaag accaaagagg gggaccttcg ggcctcttgc catcagatgt gcccagatgg
 241 gattagcttg ttggtgaggt aacggctcac caaggcgacg atccctagct ggtctgagag
 301 gatgaccagc cacactggaa ctgagacacg gtccagactc ctacgggagg cagcagtggg
 361 gaatattgca caatgggcgc aagcctgatg cagccatgcc gcgtgtatga agaaggcctt
 421 cgggttgtaa agtactttca gcggggagga aggtgttgtg gttaataacc gcagcaattg
 481 acgttacccg cagaagaagc accggctaac tccgtgccag cagccgcggt aatacggagg
 541 gtgcaagcgt taatcggaat tactgggcgt aaagcgcacg caggcggtct gtcaagtcgg
 601 atgtgaaatc cccgggctca acctgggaac tgcattcgaa actggcaggc ttgagtcttg
 661 tagagggggg tagaattcca ggtgtagcgg tgaaatgcgt agagatctgg aggaataccg
 721 gtggcgaagg cggccccctg gacaaagact gacgctcagg tgcgaaagcg tggggagcaa
 781 acaggattag ataccctggt agtccacgcc gtaaacgatg tctacttgga ggttgtgccc
 841 ttgaggcgtg gcttccggag ctaacgcgtt aagtagaccg cctggggagt acggccgcaa
 901 ggttaaaact caaatgaatt gacgggggcc cgcacaagcg gtggagcatg tggtttaatt
 961 cgatgcaacg cgaagaacct tacctggtct tgacatccac agaactttcc agagatggat
1021 tggtgccttc gggaactgtg agacaggtgc tgcatggctg tcgtcagctc gtgttgtgaa
1081 atgttgggtt aagtcccgca acgagcgcaa cccttatcct ttgttgccag cgattaggtc
1141 gggaactcaa aggagactgc cagtgataaa ctggaggaag gtggggatga cgtcaagtca
1201 tcatggccct tacgaccagg gctacacacg tgctacaatg gcgcatacaa agagaagcga
1261 cctcgcgaga gcaagcggac ctcataaagt gcgtcgtagt ccggattgga gtctgcaact
1321 cgactccatg aagtcggaat cgctagtaat cgtggatcag aatgccacgg tgaatacgtt
1381 cccgggcctt gtacacaccg cccgtcacac catgggagtg ggttgcaaaa gaagtaggta
1441 gcttaacctt cgggagggcg cttaccactt tgtgattcat gactggggtg aagtcgtaac
1501 aaggtaaccg taggggaacc tgcggttgga tcacctcctt acct
\end{verbatim}
}

This gene is a key component of the protein synthesis machinery in
cells. In fact, it is such a basic ingredient of life, that 
the same sequence, albeit with minor modifications, exists in the
genomes of all living organisms. For example, the reader is encouraged 
to compare it to the 16S rRNA gene in Escherichia coli (K12):
{\small 
\begin{verbatim}
   1 aaattgaaga gtttgatcat ggctcagatt gaacgctggc ggcaggccta acacatgcaa
  61 gtcgaacggt aacaggaaga agcttgcttc tttgctgacg agtggcggac gggtgagtaa
 121 tgtctgggaa actgcctgat ggagggggat aactactgga aacggtagct aataccgcat
 181 aacgtcgcaa gaccaaagag ggggaccttc gggcctcttg ccatcggatg tgcccagatg
 241 ggattagcta gtaggtgggg taacggctca cctaggcgac gatccctagc tggtctgaga
 301 ggatgaccag ccacactgga actgagacac ggtccagact cctacgggag gcagcagtgg
 361 ggaatattgc acaatgggcg caagcctgat gcagccatgc cgcgtgtatg aagaaggcct
 421 tcgggttgta aagtactttc agcggggagg aagggagtaa agttaatacc tttgctcatt
 481 gacgttaccc gcagaagaag caccggctaa ctccgtgcca gcagccgcgg taatacggag
 541 ggtgcaagcg ttaatcggaa ttactgggcg taaagcgcac gcaggcggtt tgttaagtca
 601 gatgtgaaat ccccgggctc aacctgggaa ctgcatctga tactggcaag cttgagtctc
 661 gtagaggggg gtagaattcc aggtgtagcg gtgaaatgcg tagagatctg gaggaatacc
 721 ggtggcgaag gcggccccct ggacgaagac tgacgctcag gtgcgaaagc gtggggagca
 781 aacaggatta gataccctgg tagtccacgc cgtaaacgat gtcgacttgg aggttgtgcc
 841 cttgaggcgt ggcttccgga gctaacgcgt taagtcgacc gcctggggag tacggccgca
 901 aggttaaaac tcaaatgaat tgacgggggc ccgcacaagc ggtggagcat gtggtttaat
 961 tcgatgcaac gcgaagaacc ttacctggtc ttgacatcca cagaactttc cagagatgga
1021 ttggtgcctt cgggaactgt gagacaggtg ctgcatggct gtcgtcagct cgtgttgtga
1081 aatgttgggt taagtcccgc aacgagcgca acccttatct tttgttgcca gcggtccggc
1141 cgggaactca aaggagactg ccagtgataa actggaggaa ggtggggatg acgtcaagtc
1201 atcatggccc ttacgaccag ggctacacac gtgctacaat ggcgcataca aagagaagcg
1261 acctcgcgag agcaagcgga cctcataaag tgcgtcgtag tccggattgg agtctgcaac
1321 tcgactccat gaagtcggaa tcgctagtaa tcgtggatca gaatgccacg gtgaatacgt
1381 tcccgggcct tgtacacacc gcccgtcaca ccatgggagt gggttgcaaa agaagtaggt
1441 agcttaacct tcgggagggc gcttaccact ttgtgattca tgactggggt gaagtcgtaa
1501 caaggtaacc gtaggggaac ctgcggttgg atcacctcct ta
\end{verbatim}
}
There are very few differences between the genes. The Salmonella sequence
has three extra bases at the end, and the E-coli gene one extra in the
beginning, there are 37 differences within the genes, and no
insertions or deletions. Such a comparison is easy to perform in the example
above, but if there are many sequences and lots of insertions and
deletions, it can be
non-trivial to identify the relationships among individual bases. This
{\em multiple sequence alignment problem} is a major problem in comparative
genomics, and is discussed in Section 3. In fact, one of our main points
is that finding a good multiple alignment is the essence of the
ancestral reconstruction problem.

The degree of conservation of the 16S rRNA gene throughout the tree of
life means it is a good starting point for reconstructing ancestral genomes. In 
particular, we can begin modestly by asking only for the ancestral 16S rRNA 
gene sequences. Such a reconstruction entails the following:
\begin{enumerate}
\item Identifying the 16S gene in many organisms, and determining the
  sequences.
\item Obtaining a likely phylogenetic tree that relates the
  sequences. 
\item Inferring ancestral sequences corresponding to internal nodes in the
  tree. 
\end{enumerate}
This comparative genomics programme was outlined by Woese et
al. \cite{Woese}. It followed on the heels of the first comparative
genomics papers, written by Linus Pauling and Emile Zuckerkandl
\cite{Pauling1,Pauling2,Pauling3}. They
applied fingerprinting techniques to compare amino acid sequences of
hemoglobins, finding that distant species have more divergent
sequences than related species. The biological problem of identifying the 16S gene and 
rapidly finding its sequence was solved in \cite{Lane}. More recently, 
new approaches have been suggested for obtaining 16S rRNA sequences, 
even from unculturable bacteria, using ``community
sequencing'' approaches \cite{Chen}.

The narrow focus of comparative genomics on 16S rRNA ended with the arrival of fast and cheap
sequencing technologies. A recent flood of ideas and research inspired
by vast amounts of genome sequences has led to
the reconstruction of numerous protein sequences and even megabases
of boreoeutheriean ancestral chromosomes. Interesting examples of the
former are \cite{Gaucher,Kuang,Thornton} and of the latter
\cite{Blanchette,Ma}. There have been a number of recent surveys on
ancestral reconstruction \cite{Rocchi,Thorntonrev}, and a  book on the topic will appear next year \cite{Liberles}.

We begin in Section 2 by discussing the ``easy'' case of ancestral
sequence reconstruction, where the phylogenetic tree is known and the alignment of the sequences is
trivial. Even in this simplest case, the choice of an effective statistical
model for evolution is non-trivial and extremely important. We introduce
the reader to these issues by way of the simplest example possible, and
provide pointers for further reading. In Section 3 we discuss
complexities that arise when inferences need to be made about insertions
and deletions, and when the alignment of the sequences is
non-trivial. The difficulty of alignment is explored in more detail in
Section 4, where we present evidence,
based on our own recent work, that suggests the amount of insertion and
deletion in genomes has been vastly underestimated. This has major 
implications for ancestral genome reconstruction. In Section 5
we discuss the problem of tree reconstruction, which needs to be solved
in the case where the phylogenetic history of the genomes being compared
is unknown. This leads us to the field of phylogenetics \cite{Steel},
where we restrict ourselves to mentioning a number of recent theoretical
advances pertinent to ancestral genome reconstruction.
We conclude in Section 6 with a
list of open problems and a discussion of the role of mathematics, 
statistics, and computer science in reconstructing ancestral genomes.

\section{Reconstructing ancestral sequences: the ``easy'' case}

In the introduction, we mentioned the example of 16S rRNA sequences, 
and observed that these genes are conserved in all organisms.
However, within restricted domains of the tree of life, there are examples
of functional elements exhibiting even more sequence conservation than
rRNA genes. The term {\em ultra-conserved elements} was introduced in
\cite{Bejerano}, and is used to described genome sequences that have
remained unchanged over millions 
of years. Such sequences were first discovered in vertebrates, and 
their degree of conservation is astounding. 
\begin{example}
\label{ex:MOL}
Consider the
sequence
\begin{equation}
\label{MOL}
{\tt tttaattgaaagaagttaattgaatgaaaatgatcaactaag}
\end{equation}
It is located in the human genome on chromosome 7, coordinates
156,694,482--156,694,523 (version March 2006). The {\em identical} sequence appears in the genomes of every other sequenced vertebrate
species to date: the chimpanzee, rhesus macaque, cat, dog, cow, mouse,
rat, rabbit, armadillo, opossum, chicken, frog,
zebrafish, pufferfish and fugufish. An alignment of the sequence
(including an extra 6 bases on each end) is shown below: 

\begin{verbatim}
     Human  tatctatttaattgaaagaagttaattgaatgaaaatgatcaactaagcttgta
     Chimp  tatctatttaattgaaagaagttaattgaatgaaaatgatcaactaagcttgta
   Macaque  tatctatttaattgaaagaagttaattgaatgaaaatgatcaactaagcttgta
       Cow  tatctatttaattgaaagaagttaattgaatgaaaatgatcaactaagcttgta
     Mouse  tatctgtttaattgaaagaagttaattgaatgaaaatgatcaactaagcttgta
       Rat  tatctgtttaattgaaagaagttaattgaatgaaaatgatcaactaagtttgta
    Rabbit  tatctatttaattgaaagaagttaattgaatgaaaatgatcaactaagcttgta
       Cat  tatctatttaattgaaagaagttaattgaatgaaaatgatcaactaagcttgta
       Dog  tatctatttaattgaaagaagttaattgaatgaaaatgatcaactaagcttgta
 Armadillo  tatctatttaattgaaagaagttaattgaatgaaaatgatcaactaagcttgta
   Opossum  tatctatttaattgaaagaagttaattgaatgaaaatgatcaactaagcttgta
   Chicken  tatctatttaattgaaagaagttaattgaatgaaaatgatcaactaagcttgta
      Frog  tatctatttaattgaaagaagttaattgaatgaaaatgatcaactaagcttgta
Pufferfish  tatctatttaattgaaagaagttaattgaatgaaaatgatcaactaagcttgta
 Zebrafish  tatctatttaattgaaagaagttaattgaatgaaaatgatcaactaagcttgta
  Fugufish  tatctatttaattgaaagaagttaattgaatgaaaatgatcaactaagcttgta
            ***** ****************************************** *****
\end{verbatim}
\end{example}
We postpone providing a precise definition of alignment until Section 3,
but remark that for our purposes here it suffices to consider an
alignment to be an ordered collection of columns. Each column
contains bases from different species that are {\em homologous}, i.e.,
that are derived from a shared common ancestral base.
The sequence in (\ref{MOL}) is called ultra-conserved,
because there appear to have been no insertions, deletions or mutations
since the common ancestor (* characters indicate columns with all
characters identical) among each group of homologous nucleotides. In
fact, in \cite{MathPhyl} we prove
\begin{theorem}
\label{thm:MOL}
The probability that the sequence (\ref{MOL}) was not present in the
genome of the ancestor of all vertebrates is less than $10^{-50}$,
assuming a Jukes-Cantor model of evolution for the sequences.
\end{theorem}
The Jukes-Cantor model is a statistical model for the evolution of
characters on trees, which is explained below. While the model has many
drawbacks and does not describe the full extent and structure of mutation,
the tiny probability is robust to changes in the model, and it is fairly
certain that (\ref{MOL}) is the ancestral sequence. 

The starting point for specifying an evolutionary model for biological
sequences is the data: a collection of $k$ sequences $\sigma^1,\ldots, \sigma^k$
of lengths $n_1,\ldots,n_k$, each with characters from a finite alphabet
$\Sigma$. We use the notation $\sigma^a_i$ to denote the $i^{th}$
element of a sequence and by a set of characters
$S=\{\sigma^1,\ldots,\sigma^k\}$ we mean the set of $n_1+ n_2
\cdots+n_k$ sequence characters that form the sequences $\sigma^1,\ldots,\sigma^k$.
These sequences may be DNA sequences (in which case the alphabet
has size $4$), amino acid sequences  (in which case the alphabet has size $20$), 
or they may be sequences derived from organisms. For example, the alphabet
may have size $2$, and the sequences may represent the presence or 
lack of genes, or morphological features in different species. 
The elements of the sequences are called {\em characters}, and 
it is important to note that in problems of interest, $k$ is small but
$n$ is usually fairly large (in the case of DNA sequences, $n$ may be in the billions). 

We restrict our discussion here to the 
evolutionary model called the {\em Cavender--Farris model} \cite{Cavender}. This is the simplest
of a class of continuous time Markov models for trees
that are used in biological sequence analysis, and although
it is very simple, it
captures many of the elements of the evolutionary models that are used
in practice. The Cavender--Farris model is an evolutionary model 
for {\em binary} sequences (i.e., the alphabet has size $2$), and 
for sequences of the same length ($n_1 = \cdots = n_k = n$).

The first ingredient of the Cavender-Farris model is a directed rooted tree 
on $k$ leaves. 
The root of the tree has biological significance: it is the 
common ancestor of all the sequences. The leaves of the tree also 
have special significance: they correspond to the $k$ different 
sequences whose evolution is being modeled. The internal vertices
of the tree correspond to speciation events. Edges of the tree
are directed from the root to the leaves, and the directions
of the arrows specify the direction of time. 

The second ingredient of the Cavender-Farris model is a 
collection of $2 \times 2$ matrices associated
with each edge of the tree. These matrices are all of the form
\begin{equation}
\label{CavFarrate}
\theta_i \quad = \quad
\begin{pmatrix}
\mu_i & \pi_i \cr 
\pi_i & \mu_i
\end{pmatrix}.
\end{equation}
where $i$ is an edge of the tree, $0 \leq \pi_i \leq 1$, and $\mu_i = 1 - \pi_i$. 
These matrices have a biological interpretation: they describe the
probabilities of characters changing along branches of the tree.
Although in principle these probabilities may be different 
for each edge {\em and} for each character within each sequence, 
the Cavender--Farris model specifies a single probability of change $\pi_i$ for 
each edge $i$. If the vertices adjacent to the edge $i$ are $v_i$ and 
$w_i$ with $i$ oriented from $v_i$ to $w_i$, then for the $j$th
character of the sequence at $v_i$, $\pi_i$ is the 
probability that the $j$th character of the sequence at $w_i$ is different. 

The Cavender-Farris model is an example of a {\em continuous time Markov
  chain model on a tree}. To see the connection to Markov models,
consider the $2 \times 2$ square matrix
\begin{equation}
\label{Qmatform}
Q \quad = \quad
\begin{pmatrix}
- \alpha & \alpha\cr 
\alpha & -\alpha 
\end{pmatrix}, \, \alpha \geq 0.
\end{equation}
The rows and columns of $Q$ are 
indexed by $\Sigma = \{{\tt 0,1}\}$.
Note that the matrix $Q$ has the following properties:
\[ q_{ij} \geq 0 \quad \,\, \hbox{for} \,\,\, \ i \neq j, \]
\[ \sum_{j \in \Sigma} q_{ij} =0 \quad \,\,\hbox{for all} \,\, \ i \in
\Sigma, \]
\[ q_{ii}  < 0 \quad \hbox{for all} \,\, \ i \in \Sigma. \]
Any square matrix $Q$ (of arbitrary size) with the above properties is called a 
{\em rate matrix}. The following is a straightforward 
result about continuous time Markov chains \cite{ASCB2005}.
\begin{theorem}
Let $Q$ be any rate matrix and $\,\theta(t)=e^{Qt} = \sum_{i=0}^\infty
\frac{1}{i \, !} Q^i t^i $. Then
\begin{enumerate}
\item $ \theta(s+t)\,=\,\theta(s) \cdot \theta(t) \qquad $ (Chapman--Kolmogorov \index{Chapman--Kolmogorov equations} equations),
\item $ \theta(t)\,$ is the unique solution to the forward
differential equation \hfill \break $\, \theta'(t)=\theta(t) \cdot Q, \,\theta(0)=
  {\bf 1} \,$ for $\,t \geq 0$ (here ${\bf 1}$ is the identity
  matrix),
\item $\theta(t)\,$ is the unique solution to the
backward differential equation \hfill \break $\, \theta'(t) = Q \cdot
\theta(t),\, \theta(0)= {\bf 1} \,$ for $\,t \geq 0$,
\item $\theta^{(k)}(0) = Q^k$.
\end{enumerate}
Furthermore, a matrix $Q$ is a rate matrix if and only if the matrix
$\theta(t) = e^{Qt}$ is a stochastic matrix (non-negative with row sums
equal to one) for every $t \geq 0$.
\end{theorem}

Note that for the binary rate matrix (\ref{Qmatform}), we have
\[ \theta(t) \quad = \quad \frac{1}{2}
\begin{pmatrix}
1 + e^{-2\alpha t_i} & 1 - e^{-2\alpha t_i}  \cr
1 -  e^{-2\alpha t_i } & 1 + e^{-2\alpha t_i}
\end{pmatrix}  . \]
The expected number of mutations over time $t$ is the quantity
\begin{equation}
\label{branchlength} \alpha t \quad = \quad -\frac{1}{2} \cdot {\rm
trace}(Q) \cdot t \quad = \quad - \frac{1}{2} \cdot {\rm log} \,{\rm
  det} \bigl( \theta(t) \bigr).  \end{equation} This number is called the
  {\em branch length}. \index{branch length} It can be computed from
  the substitution matrix $\theta(t)$ and is the expected number of
  mutations. For this reason it is used instead of $t$ to label edges in a phylogenetic
  tree.

The matrices $\theta_i$ in the Cavender-Farris model are parameterized
by $\alpha$ and $t$:
\begin{equation}
\mu_i = \frac{1}{2}(1+e^{-2\alpha t}), \quad \pi_i =
\frac{1}{2}(1-e^{-2\alpha t}).
\end{equation}

One way to specify an evolutionary model is to give a phylogenetic tree $T$ together with $Q$ and an
initial distribution for the root of $T$ (which we here assume to be the
uniform distribution on $\Sigma$).  The branch lengths of the edges are
unknown parameters, and the objective is to estimate these branch
lengths from data. Thus, if the tree $T$ has $r$ edges, then such a model has $r$ free parameters.

Returning to Theorem \ref{thm:MOL}, we note that the Jukes-Cantor model
is just the Cavender-Farris model with $|\Sigma|=4$. That is, the $Q$
matrix is given by
\begin{equation}
\label{QmatformJC}
Q \quad = \quad
\begin{pmatrix}
-3\alpha & \alpha & \alpha & \alpha \cr 
\alpha & -3\alpha & \alpha & \alpha \cr
\alpha & \alpha & -3\alpha & \alpha \cr
\alpha & \alpha & \alpha & -3\alpha 
\end{pmatrix}, \, \alpha \geq 0.
\end{equation}
In this case the branch length is given by $3 \alpha t$ (this should be
compared with (2.4)).

The Cavender-Farris/Jukes-Cantor models are too simple to be used in
practice. Point mutations in genome display various asymmetries, and the
general reversible Markov model is preferred. The model in (2.7,2.8)
is realistic, and was estimated from observed synonymous substitutions (those that
do not change the amino acid) in human-mouse-rat alignments
\cite{Yap}. Note that $\pi_a,\pi_C,\pi_G$ and $\pi_T$ are the
equilibrium frequencies, and are also parameters in the model.
\begin{eqnarray}
\label{Q-matrix}
&  Q \quad = \quad
\begin{pmatrix}
-1.05 & 0.19 & 0.71 & 0.15 \cr
 0.17 & -0.96 & 0.18 & 0.61\cr
 0.60 & 0.17 & -0.95 & 0.17\cr
 0.15 & 0.72 & 0.20 & -1.07
\end{pmatrix}, &\\ 
 & \pi_A=\pi_T = 0.23, \pi_G = \pi_C = 0.27. &
\end{eqnarray}
The models used can be even more general,
including local dependencies between sites and different
functional categories for the ancestral sequences that alter mutation
rates. There is an extensive literature on this topic,
as well as many papers discussing the reliability of reconstructed characters
in aligned sequences (e.g. \cite{Pupko}).

The final aspect of Theorem \ref{thm:MOL} that we have not yet discussed
is a computational one, namely how to compute the probability given
the sequences, the tree and the model. The algorithm is known as
Felsenstein's algorithm \cite{Fel81} and involves dynamic programming
on a tree. In the context of 
ancestral reconstruction, \cite{PupkoFast} show how to modify
Felsenstein's algorithm for fast joint reconstruction of all ancestral
sequences. It is best to view all these algorithms
as a special case of the Junction Tree algorithm for a graphical model
where the graph is a tree.

We conclude by noting that there is a direct connection between evolutionary models such as the
Cavender-Farris model and the emerging field of {\em algebraic statistics}. This is
because the families of probability distributions on the leaves 
are essentially parameterized algebraic varieties, and for this reason the tools
of commutative algebra and algebraic geometry can be used to study the model and develop inference
methods. We refer the interested reader to the recent book
\cite{ASCB2005} for an introduction to the subject.

\section{Alignment}

The models for ancestral reconstruction in Section 2 do
not account for
insertions and deletions (indels). This is a serious drawback because
the homology between muliple sequences is complicated by insertions,
deletions, rearrangements, segmental duplications, and other evolutionary
events. We restrict our discussion in this section to the issues
regarding ancestral reconstruction in the presence of insertions and
deletions only.

\begin{definition}
\label{def:align}
A {\em partial global multiple alignment} of sequence characters $S=\{\sigma^1,\ldots,\sigma^k\}$
is a partially ordered set $P=\{c_1,\ldots,c_m\}$ together 
with a surjective function $\varphi:S \rightarrow P$ such that 
$\varphi(\sigma^a_i) < \varphi(\sigma^a_j)$ if $i< j$.
\end{definition}
The elements of $P$ correspond to columns of the multiple alignment, 
and the partial order specifies the order in which columns must appear. 
Note that the function $\varphi$ specifies the homology relationships
among the sequences: two characters that are mapped to the same element
in the poset are homologous. Thus, an alignment is just a pair $(P,\varphi)$.
We call $P$ an {\em alignment poset}, and 
note that unless $P$ is a
total order (chain), there are columns of the partial 
multiple alignment whose order is unspecified. 
\begin{figure}[ht]
\includegraphics[scale=0.15]{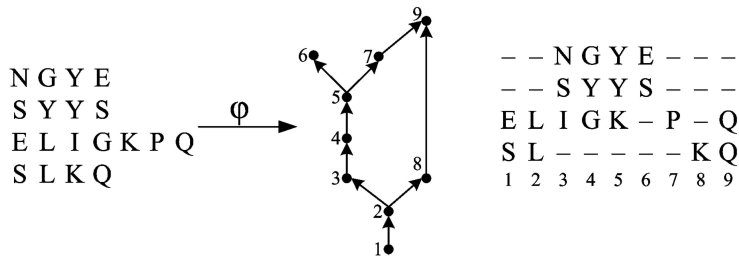}
\caption{A set of four sequences, an alignment poset together
with a linear extension, and a global multiple alignment. The function 
from the set of sequence elements to the alignment poset that 
specifies the multiple alignment is not shown, but is fully 
specified by the diagram on the right. For example, the second
element in the first sequence is $\sigma^1_2=G$, 
and $\varphi(\sigma^1_2)$ corresponds to the fourth column of the multiple alignment.
}\label{fig:01}
\end{figure}
A {\em linear extension} of a partially ordered set $P=\{c_1,\ldots,c_m\}$ 
is a permutation of the elements $c_1,\ldots,c_m$ such that
whenever $c_i < c_j$, $i<j$. A {\em global multiple alignment} is a 
partial global multiple alignment together with a linear extension of 
the alignment poset $P$ (see Figure \ref{fig:01}). The alignment in
example \ref{ex:MOL} is a global multiple alignment where the poset is a chain.

The problem of finding a multiple alignment for a collection of
sequences is to find a ``good'' poset and function $\varphi$ that are
likely to determine the homology correctly. This is a non-trivial
problem, and its solution is
essential for accurate ancestral reconstruction of sequences. A complete
discussion of multiple alignment is beyond the scope of this section,
but we briefly review pairwise sequence alignment.

In the case of two sequences, an alignment poset is always a disjoint
union of chains. Note that for two sequences of length $n$ and $m$,
there are ${n + m \choose n}$ alignments. We consider a simple criterion
for selecting an alignment. Given three parameters $X,S$ and
$M$, we assign a score to an alignment as follows:
\begin{equation}
\label{eq:alignment} {\rm score} \, (P,\varphi) \quad = \quad M \cdot \#M + X \cdot \#X +
S \cdot \#S, 
\end{equation}
where $\#S = |\{x \in P:|\varphi^{-1}(x)| =1 \}|$ (the number of spaces
in the alignment), $\#M = |\{a \in \sigma^i, b \in
\sigma^j:\varphi(a)=\varphi(b), \, {\rm char} \,  a= {\rm char} \, b \}|$ (the number of
matches), and $\#X = |\{a \in  \sigma^i, b \in
\sigma^j:\varphi(a)=\varphi(b), \, {\rm char} \, a \neq {\rm char} \, b\}|$ (the number of mismatches). The problem of maximizing 
(\ref{eq:alignment}) can be solved efficiently using dynamic
programming, using what is known as the {\em Needleman-Wunsch algorithm}
\cite{Needleman}. As with Felsenstein's algorithm, this is
just a special case of the Junction Tree algorithm, in this case for a
graphical model known as the {\em pair hidden Markov model}. Despite the
name, pair hidden Markov models
are non-trivial modifications of standard hidden Markov models
models, in that the structure of the models is not fixed. In the
graphical model literature, these types of models are referred to as
Bayesian multinets.
The connection between alignments and pair HMMs provides a probabilistic
interpretation for the score in (\ref{eq:alignment}). For details see
\cite{PairHMM} or \cite{ASCB2005}.

An alignment specifies the homology relationships among nucleotides or
amino acids, but provides no information about ancestral sequences. In
order to make inferences about ancestral sequences, it is necessary to
use the alignment to estimate the locations, sizes, and times of insertions
and deletions. 

\begin{figure}[ht]
\begin{center}
\includegraphics[scale=0.37]{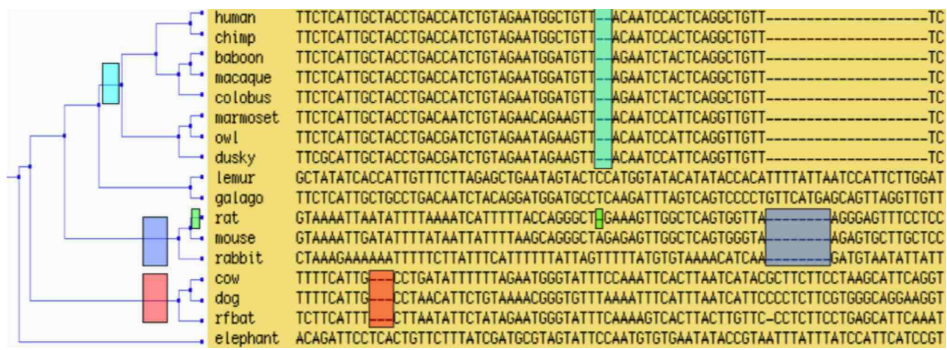}
\caption{Alignment of 17 species from the CFTR region \cite{Snir}.}
\end{center}
\end{figure}

\begin{example}
Figure 2 shows an alignment of 17 species from the cystic fibrosis transmembrane
  conductance regulatory region (CFTR) produced for the
  ENCODE consortium \cite{ENCODE}. The phylogenetic tree on the left can
  be used in conjunction with the alignment to identify indel
  events. The boxes shown in the figure show the most parsimonious
  explanation (minimum number of events), based on the tree \cite{Snir}. 
\end{example}

Given a phylogenetic tree $T$, together with a pair hidden Markov model
associated to each edge of $T$, the {\em tree alignment problem} is to 
find sequences $\sigma^i$ where $i$
ranges over all the internal nodes of $T$ such that
$\prod_{e=(i,j) \in E(T)} P(\sigma^i,\sigma^j)$ is maximized. The
inference performed in Example 3.2 \cite{Snir} is based on a restricted
version of tree alignment. In general, there is a 
ratio-two approximation algorithm for the problem that runs in
quadratic time, together with a polynomial time approximation scheme \cite{Lusheng}.

\section{The limits of ancestral reconstruction: indel saturation}

Studies of the reliability of ancestral reconstruction have mostly
focused on point
mutations and the effects of using different types of evolutionary models
\cite{Williams}. Such studies implicitly assume that multiple alignment,
while difficult, is a tractable problem and that indels can be
effectively ignored, or else accounted for using procedures such as tree
alignment. In this
section we provide indirect evidence that even over relatively short time-scales,
large number of insertions and deletions make it impossible to align
sequences (as there is little or no homology), and therefore it is
impossible to reconstruct ancestral sequences. Our arguments are based
on estimating the overall numbers of indels from length differences of
homologous segments. 

We illustrate our ideas with a calculation based on the comparison of
introns in four species of the Drosophila (fruit fly): melanogaster, pseudoobscura,
yakuba and virils. Introns are sequences within genes that are spliced out
after transcription but before translation, so that they do not
contribute to the amino acids in the protein. Although it has been shown
that introns can be important for regulating the expression of genes,
they consist of mostly non-functional sequence. 
The reason to examine
Drosophila is that multiple genomes have been sequenced during the past
year. The reason to look at introns is that it is easy to identify
homologous introns when the genes they reside in are unambiguously homologous.

Our methods provide an
overall assessment of the number of inserted and deleted nucleotides
along each branch of the tree relating the species. The inference is
based on a maximum likelihood model for Brownian motion on a tree,
originally developed by Felsenstein \cite{Felsenstein1981b} for modeling
changes in gene frequencies over time. Our application of this model to
estimating indel rates from length differences of introns is new, and is
the first such analysis with more than a pair of species.
Previous work on the comparison of lengths of pairs of homologous
introns \cite{Ogurtsov,Yandell} has mainly revealed that there is a correlation
between the length differences and the distance between species.

We explain the Gaussian model we consider completely but briefly, for more
detail and background we refer the reader to \cite{Felsenstein1981b,Pachterindel}.
Let $T$ be an unrooted tree with $n$ leaves, together with labels
${\bf v} = v_1,\ldots,v_{2n-3}$ on the edges parameterizing variances of normal
distributions. 
Let $x^s_{ij}$ be quantitative characters generated from a Brownian
motion model on a tree.
We use the term quantitative character to refer to uncountable state
spaces, in this case $x^s_{ij} \in \mathbb{R}$. The $ij$th contrast is
$\Delta x^s_{ij} = x^s_i - x^s_j$. Note that $\Delta x^s_{ij}$ is drawn
from a normal distribution with mean $0$ and variance $v_{ij}$.
Let $C=\{i_1j_1,\ldots,i_{n-1}j_{n-1}\} \subset {[n] \choose
  2}$ be a spanning tree of the complete graph $K_n$.
In the case of four taxa which we will consider (see Figure 3), the $C$-covariance matrix
for $C=\{12,13,14\}$ is 
\begin{equation}
\label{eq:contrastmatrix}
 \Sigma_{12,13,14} = \left(\begin{matrix}
 v_{15}+v_{25} & v_{15}         & v_{15}         \\
 v_{15}     & v_{15}+v_{36}+v_{56} & v_{15}+v_{56}     \\
 v_{15}     & v_{15}+v_{56}     & v_{15}+v_{46}+v_{56} \\
\end{matrix}  \right).
\end{equation}

We let $\Delta {\bf x}_C^s = [\Delta x^s_{i_1j_1},\ldots,\Delta
  x^s_{i_{n-1}j_{n-1}}]$ and 
denote the determinant of $\Sigma_C$ by $|\Sigma_C|$. The log-likelihood
of the data is given by
\begin{equation}
\label{eq:loglik}
\mbox{ln}\, \tilde{L}(\Delta {\bf x}_{{[n] \choose 2}}^s|{\bf v}) = -\frac{(n-1)p}{2}\mbox{ln}\, 2\pi - \frac{p}{2}ln\, |\Sigma_{C}|+\frac{1}{2}
\sum_{s=1}^{p} \left( (\Delta{\bf  x}_C^s)^T \Sigma_C^{-1} \Delta {\bf  x}_C^s \right).
\end{equation}
This specifies the model we will use completely. The next Lemma follows from the ``Pulley principle'' \cite{Felsenstein1981b}.
\begin{lemma}
\label{lem:dist}
The log-likelihood (\ref{eq:loglik}) does not depend on the choice of
$C$. Furthermore, (\ref{eq:loglik}) is linear in the ${n \choose
  2}$ numbers $d_{ij} = \sum_{s=1}^{p} \left( \Delta x^s_{ij}
\right)^2$.
\end{lemma}
In other words, if we let ${\bf d} = \{d_{ij}\}_{i,j=1}^{n}$, then the maximum likelihood estimator
\begin{equation}
\label{eq:likvd}
\widehat{{\bf v}_{d}} = \mbox{argmax}_{\bf v} \quad \mbox{ln} L({\bf d}|{\bf v}).
\end{equation}
is well-defined and it therefore makes sense to refer to the ``log-likelihood function for the contrast Brownian motion model
on a tree''.
\begin{example}[n=3, from \cite{Felsenstein1981b}]
The likelihood equation becomes
\begin{eqnarray*}
\mbox{ln}\, L(d_{12},d_{13,},d_{23}|v_{14},v_{24},v_{34}) & = &
-p\mbox{ln}\, 2\pi - \frac{p}{2}\mbox{ln}\,
(v_{14}v_{24}+v_{14}v_{34}+v_{24}v_{34})\\
& & - \frac{v_{14}d_{23}+v_{24}d_{13}+v_{34}d_{12}}{2(v_{14}v_{24}+v_{14}v_{34}+v_{24}v_{34})}.
\end{eqnarray*}
The critical equations are easy to solve and one finds that
\begin{eqnarray*}
\widehat{v_{14}} &  = & (d_{12}+d_{13}-d_{23})/(2p),\\
\widehat{v_{24}} & =  & (d_{23}+d_{12}-d_{13})/(2p),\\
\widehat{v_{34}} & =  & (d_{13}+d_{23}-d_{12})/(2p).
\end{eqnarray*}
\end{example}
Note that in the case $p=1$, if ${\bf d}$ is a tree metric then $\widehat{V_{ij}} = d_{ij}$. This is true
for general $n$, and is a restatement of the fact that the maximum likelihood estimator
is consistent.

\begin{figure}
\begin{center}
\includegraphics[scale=0.4]{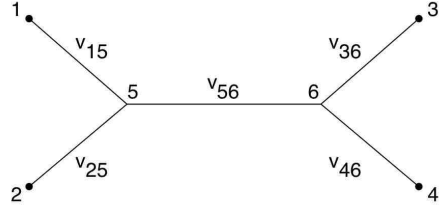}
\caption{Tree with four leaves.}
\end{center}
\end{figure}

\begin{example}[n=4] There are five critical equations in the $n=4$ case:
\begin{eqnarray*}
v_{25}(v_{36}+v_{46})+v_{56}(v_{36}+v_{46}) & = & 
\left( d_{34}(v_{25}+v_{56}) + d_{24}v_{35} + d_{23}v_{45}
\right)/(2p), \\
v_{15}(v_{36}+v_{46})+v_{56}(v_{36}+v_{46}) & = & 
\left( d_{34}(v_{15}+v_{56}) + d_{13}v_{46} + d_{14}v_{36}
\right)/(2p), \\
v_{46}(v_{36}+v_{46})+v_{56}(v_{14}+v_{24}) & = & 
\left( d_{12}(v_{46}+v_{56}) + d_{14}v_{25} + d_{24}v_{15}
\right)/(2p),\\
v_{36}(v_{36}+v_{46})+v_{56}(v_{14}+v_{24}) & = & 
\left( d_{12}(v_{36}+v_{56}) + d_{23}v_{15} + d_{13}v_{25}
\right)/(2p), \\
(v_{15}+v_{25})(v_{36}+v_{46})  & = &   \left(
d_{34}(v_{15}+v_{25}) + d_{12}(v_{36}+v_{46}) \right)/(2p).
\end{eqnarray*}
The solution of these equations is an exercise in elimination, and can
be done using Gr\"{o}bner bases methods \cite{ASCB2005}.
Using the pulley principle we can restrict ourselves to $\widehat{v_{15}}+\widehat{v_{25}} = d_{12}$ and 
$\widehat{v_{36}}+\widehat{v_{46}} = d_{34}$, in which case we find one unique
critical point: the global maximum. The solution consists of 5
enormous rational functions in the $d_{ij}$ which we omit here due to lack of
space.
\end{example}

\begin{figure}
\begin{center}
\includegraphics[scale=0.25]{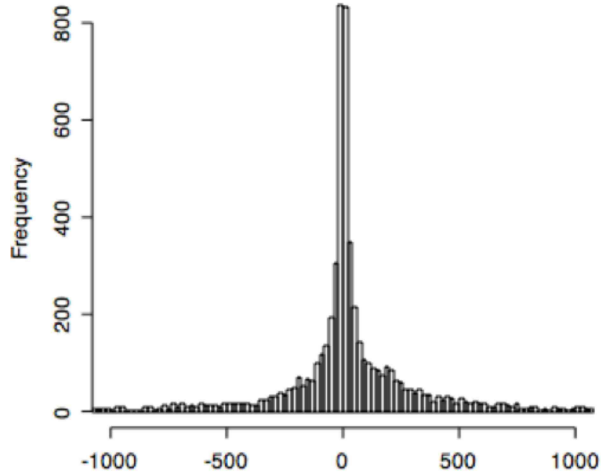}
\caption{Differences in lengths between homologous introns in
Drosophila melanogaster and pseudoobscura. Introns with similar numbers of insertions and deletions
display little length difference, but that does not mean that introns
with small length differences have not undergone insertion and deletion.}
\end{center}
\end{figure}

The data analyzed was obtained from \cite{Yandell}. 
We restricted our attention
to introns in the range of $100-500$bp, and cleaned up the dataset by
removing duplicates or cases with ambiguous homology.
We then computed, for each pair of species, the ``distance''
\begin{equation}
d_{ij}=\sum_{s=1}^{p}(\Delta x^s_{ij})^2,
\end{equation}
 where $\Delta x^s_{ij}$ is the
difference in length between the $s$th pair of introns in species $i$ 
and $j$. This is just the quantity that appears in Lemma
\ref{lem:dist}. It is important to note that the use of the Brownian
motion model on the $d_{ij}$ constitutes an approximation to a Poisson
process model for indels. We omit the details of this relationship and
again refer to \cite{Pachterindel}. 

Returning to the data, we show a histogram of the differences between
the intron lengths for a pair of species in Figure 4. These are the raw
data we use to compute the $d_{ij}$ distances, and then the maximum
likelihood estimates for the branch lengths, which are the total number
of indels. The maximum likelihood estimates are 
\begin{eqnarray*}
\widehat{v}_{dmel,5} & = & 3204,\\
\widehat{v}_{dyak,5} & = & 2521,\\
\widehat{v}_{dpse,6} & = & 9595,\\
\widehat{v}_{dvir,6} & = & 49894,\\
\widehat{v}_{56} & = & 16169,\\
\end{eqnarray*}
where vertices $5,6$ are the internal vertices in the four taxa tree (as
in Figure 3). In other words, we estimate that the total number of
inserted and deleted bases between D. melanogaster and D. Yakuba is
5725, and between D. melanogaster and D. pseudoobscura we obtain $3204+16169+9595=28968$.

These numbers are much larger than the mean length of the introns we
considered (note that the longest introns had length $500$). The
conclusion is that the {\em total} number of inserted and deleted bases
is far larger than the size of the intron. This may seem surprising at
first, but is a reflection of the fact that insertions and deletions
cancel each other out, and therefore a small difference in intron length
does not necessarily indicate a lack of indel activity. In fact, the
intuition that homologous introns of similar length must contain
homologous nucleotides is false. The difference of two Poisson
distributions is Skellam distributed, and if
the rates are the same then the peak is at 0, exactly what we see in
Figure 4.

The Brownian motion model we have proposed is (too) simple; it is the
indel equivalent of the Jukes-Cantor model for point mutations, but it
is sufficiently realistic to suggest that it may be impossible to
reconstruct ancestral intron due to excessive indel turnover. The large amount of insertion and
deletion should not be surprising in the light of the existence of {\em transposable elements}.
These are repetitive elements that make up a large
fraction of many genomes. The term transposable 
elements groups several subclasses of elements that 
replicate autonomously in the genome, either through
reverse transcription, or directly from DNA to DNA 
via excision and repair. Up to half of the human genome 
is composed of such elements, and although they are sometimes
thought of as ``parasitic elements'', somewhat like viruses, 
they clearly play an integral role in shaping genome evolution, and 
in many cases are believed to influence gene function. Unfortunately,
they confound attempts at reconstructing genomes, by virtue of creating
enormous turnover in the sequences. At the very least, a complete
catalog of such elements will be essential for reconstructing ancestral
genomes.

\section{Tree reconstruction}

In the previous sections we have been assuming that the phylogenetic
tree for the species under consideration is known. This assumption is,
unfortunately, rarely justifiable. Molecular based phylogenies may not
conclusively determine certain branchings in a tree, and fossil-based
phylogenies tend to have low resolution. We mention two best-case
examples where despite substantial work there is still some disagreement as to
the actual phylogeny. In vertebrates, molecular techniques are not in
agreement with other methods used for the rodents (the so-called ``rodent
problem'' \cite{Aidroos,Thomas}), and in Drosophila there is disagreement
about the splits among Drosophila erecta, yakuba and melanogaster
\cite{Pollard}. In other branches of the tree of life, the situation can be that nothing at all is
known about the details of the phylogeny. Thus, phylogenetic trees must
be inferred, and the topology of the trees has a direct bearing on the
reconstructed ancestral sequences \cite{Ross}.

We begin by describing a likelihood-based strategy that can be followed, but that is computationally infeasible in practice:
for each tree, 
the probability of the known sequences may be computed for a 
specific evolutionary model, and one can select the 
tree/evolutionary model combination with maximal likelihood.
This is known as the {\em maximum likelihood approach} to 
phylogeny reconstruction.
The reason the algorithm proposed above is computationally intractable
for trees with many leaves is that the number of binary
trees with $k$ leaves is $(2k-5)!!$. Moreover, the problem of finding
the maximum likelihood branch lengths for a fixed tree is very difficult \cite{Serkan}. 
The field of phylogenetics research
is very active and it is only recently that the following result was
published, quantifying the difficulty of the tree reconstruction problem:

\begin{theorem}[\cite{Chor}]
Given a set of binary strings, all of the same length, and a negative
number $L$, it is NP-hard to determine whether there is a tree $T$ such
that the log likelihood of the sequences for the tree $T$ with optimal
branch lengths is greater than $L$. 
\end{theorem}

On the positive side, there are theoretically sound approaches to
phylogenetic reconstruction that are also practical for large
datasets. In the context of ancestral reconstruction there are often
many taxa to be considered, and the favored approach is {\em neighbor
  joining} \cite{NJ} (sometimes other closely related distance-based
algorithms are used \cite{Cai}). The neighbor joining algorithm takes as
input a {\em dissimilarity map} on a set of taxa $X$. This is a map
$\delta:X \times X \rightarrow {\mathbb R}$ that satisfies $\delta(i,j)=\delta(j,i)$
  and $\delta(i,i)=0$. The quantities $\delta(i,j)$ are maximum
  likelihood estimates of the branch length (see \ref{branchlength}) between every pair of taxa. The
  algorithm is:

\begin{enumerate}
\item  Given a dissimilarity map $\delta$, compute the {\em Q-criterion}
\[
 Q_{\delta}(i,j)= (n-2) \delta(i,j) - \sum_{k \neq i} \delta(i,k) -\sum_{k \neq j}
\delta(j,k).
\]
Then select a pair $a,b$ that minimize $Q_{\delta}$ as motivated
by the following theorem:
\begin{theorem}[\cite{NJ}]
\label{thm:Saitou} Let $\delta_T$ be the tree metric corresponding
to the tree $T$. The pair $a,b$ that minimizes $Q_{\delta_T}(i,j)$
is a cherry in the tree.
\end{theorem}
\item If there are more than three taxa, replace the putative
cherry $a$ and $b$ with a leaf $j_{ab}$, and construct a new
dissimilarity map where $\delta(i,j_{ab}) =
\frac{1}{2}(\delta(i,a)+\delta(i,b))$. This is called the {\em
reduction step}. \item Repeat (1) and (2) until there are three taxa.
\end{enumerate}

Neighbor-joining is fast: existing
implementations run in $O(n^3)$ where $n$ is the number of taxa, and it
has been observed (empirically) to produce good results
\cite{Kumar}. However, despite the ubiquitous use of the algorithm, very little about
it has been understood until recently. Exciting new results include:
\begin{itemize}
\item The development of fast neighbor-joining which achieves an optimal
  run time of $O(n^2)$ \cite{Elias}.
\item A uniqueness theorem for the algorithm \cite{Bryant}.
\item An answer to the question of ``what does neighbor-joining
  optimize''? \cite{GascuelSteel}.
\item An answer to the question of ``when (and why) does
  neighbor-joining work''? \cite{MihaescuPachter}.

\end{itemize}
Together, these results provide new insight into the algorithm, and open
up the possibility for significant improvements in accuracy.

Returning to maximum likelihood phylogenetic reconstruction, 
recent results also show that it is possible to
efficiently reconstruct the topology of trees (with high probability)
using likelihood models of the type described in Section 2 given only
polylogarithmic quantities of data.

\begin{theorem}[\cite{Daskalakis}]
\label{Daskalakis}
Under the Cavender-Farris model, there is a constructive algorithm 
that reconstructs almost all trees on $k$ leaves with sequences of
length $k=O(poly({\rm log} k))$.
\end{theorem}

A recent related result \cite{Mihaescu} provides an alternative analysis 
that quantitatively couples the reconstruction problem with the
ancestral reconstruction problem.
Indeed, it appears that ancestral sequence reconstruction
and tree reconstruction are far more related than originally thought.

\section{Open problems and discussion}

A recent survey article \cite{Rocchi} proposes that ``an integrated,
multi-disciplinary approach is needed in order to make progress on
ancestral genome reconstruction''. We agree with this point of view, and
in the spirit of the proposal offer an invitation to mathematicians,
statisticians and computer scientists by highlighting some open problems
that may form a starting point for research and collaboration.  We focus on
problems important for biology, but many of the questions also lead to
interesting mathematics \cite{BerndClay}.

The Cavender-Farris model introduced in Section 2 is the simplest
example of an evolutionary model. A central problem in genomics is to
find appropriate models that effectively capture the mechanisms by which
sequences changes, but that are also useful for inference. An important
class of models that have been proposed are phylogenetic hidden Markov
models \cite{McAuliffe,Siepel}.
\begin{problem}[Phylogenetic hidden Markov models]
Find efficient algorithms for inference with phylogenetic hidden Markov models.
\end{problem}
See \cite{Jojic} for an introduction to this problem and some first
steps exploring the use of variational methods and other graphical model techniques.

In Section 4, we raise the issue of alignability of sequences, and the
implications for ancestral genome reconstruction. There are other
approaches to addressing the problem of alignability: 
In \cite{Dewey}, we show that the choice of parameters is crucial for
correctly identifying homologous transcription factor binding sites. The 
methods used are those of {\em parametric alignment}, which is a
geometric approach to studying the dependence of optimal alignments on
parameters. We propose the following problem based on \cite{Fernandez}.
\begin{problem}[Parametric ancestral reconstruction]
Develop polyhedral algorithms for the ancestral
reconstruction problem. In particular, what implications does the {\em
  Jukes-Cantor function} of \cite{Dewey}, have for ancestral reconstruction?
\end{problem}
In a similar vein, and inspired by our observation in Section 4 that the
number of indels in introns may preclude ancestral reconstruction, we ask:
\begin{problem}[Indel saturation]
What are the limits on ancestral reconstruction as determined by indel
rates and distances?
\end{problem}

In the field of phylogenetics, we offer two problems chosen for their
specific relevance to the ancestral genome reconstruction problem. For
readers interested in algebraic geometry, we mention \cite{Eriksson} for
further mathematical problems.
\begin{problem}[Tree reconstruction and alignment]
Find efficient algorithms for reconstructing a tree under the tree
alignment model.
\end{problem}
The next problem is especially important for ancestral reconstruction of
bacterial genomes where there is a lot of horizontal transfer:
\begin{problem}[Consistency theorems for networks]
Extend the robustness analysis of \cite{MihaescuPachter} to
generalizations of the neighbor joining algorithm that project dissimilarity maps onto
phylogenetic networks \cite{BryantMoulton,Huynh} rather than trees.
\end{problem}

The reconstruction of ancestral genomes involves more than the
inference of ancestral sequences based on groups of homologous extant
nucleotides. The order of sequences is also important, and an important
component of {\em whole} genome reconstruction is the inference of the
ancestral order of genomic segments. 
The problem is closely related to the whole genome alignment
problem \cite{DeweyPach}. In this regard, Definition \ref{def:align} is
too restrictive. For example, it does not allow for homology relationships where
there have been rearrangements, inversions, and segmental
duplications. In our opinion, a major problem that needs to be solved,
where a close collaboration between biologists and mathematicians is
necessary is:

\begin{problem}[What is an alignment?]
Provide a definition for  whole genome alignment that
is based on a comprehensive biological definition of homology.
\end{problem}

There are formulations of alignment different from \ref{def:align}, but
they are also too restrictive. Nevertheless, we mention one important
approach to inference of ancestral order:
\begin{definition}
A {\em reversal} alignment between two genomes is a signed permutation.
\end{definition}
For example, the signed permutation 
\begin{equation}
\label{mouseX}
1 \, 7^- \, 6 \, 10^- \, 9 \, 8^- \, 2 \, 11^-
\, 3^- \, 5 \, 4
\end{equation} is a reversal alignment between the human and mouse X
chromosomes (this is an example from \cite{Pevzner}). This means that there is a division of the
human X chromosome into $11$ pieces (equivalently 11 breakpoints) such that if they are labeled, in
order, $1 \, 2 \, 3 \, 4 \, 5 \, 6 \, 7 \, 8 \, 9 \, 10 \, 11$, then
they appear in the mouse in the order (\ref{mouseX}). Note that the
negative signs specify the direction of the segments, with a negative
indicating reversal and complementation. A reversal operation involves 
reversing the order of a segment of a signed permutation, and flipping the
signs. For example, by a reversal of (\ref{mouseX}) can consist of
changing the segment $11^- \, 3^- \, 5 \, 4$ to $4^- \, 5^- \, 3 \,
11$. Biologically, reversals correspond to rearrangement events. The
reconstruction of ancestral order is equivalent to

\begin{problem}[The median problem]
Given a phylogenetic tree $T$, a distance measure between signed
permutations, and signed permutations labeling the
leaves of $T$, find signed permutations $\pi^i$ where $i$
ranges over all the internal nodes of $T$ such that
the sum of the distances between permutations adjacent in the
tree is minimized.
\end{problem}
The case where the distance measure is the reversal distance is already
interesting and difficult, but in practice more complex distance
measures need to be used (that allow for multiple chromosomes and other
events, such as duplications). For more on the problem see
\cite{Bourque,Durrett,Wang}.

We conclude by noting that although we have not discussed it in this
paper, ancestral reconstructions of proteins can be sequenced and tested
for their physiochemical properties (e.g., \cite{Kuang,Thornton}). Thus,
ancestral reconstructions are not merely theoretical
exercises. This exciting aspect of the field continues to be developed,
and will hopefully lead to tests not just of genes, but also of
ancestral regulatory elements and larger genome segments.

\newpage
\bibliographystyle{amsalpha}

\begin{thebibliography}{A}

\bibitem{PairHMM} M. Alexandersson, N. Bray and L. Pachter, \textit{Pair
  hidden Markov models}, special review in the Encyclopedia of Genetics,
  Genomics, Proteomics and Bioinformatics (L.B. Jorde, P. Little,
  M. Dunn and S. Subramanian, eds.), 2005.

\bibitem{Aidroos} J. Al-Aidroos and S. Snir, \textit{Analysis of point
  mutations in vertebrate genomes}, in Algebraic Statistics for
  Computational Biology (L. Pachter and B. Sturmfels eds.), Cambridge
  University Press, 2006.

\bibitem{Bejerano} G. Bejerano, M. Pheasant, I. Makunun, S. Stephen,
  W.J. Kent, J.S. Mattick and D. Haussler, \textit{Ultraconserved
    elements in the human genome}, Science 304:1321--1325, 2004.

\bibitem{Blanchette} M. Blanchette, E.D. Green, W. Miller and
  D. Haussler, \textit{Reconstructing large regions of an ancestral
    mammalian genome in silico}, Genome Research 14:2412--2423, 2004.

\bibitem{Pevzner} G. Bourque, P.A.Pevzner and G. Tesler,
  \textit{Reconstructing the genomic architecture of ancestral mammals:
    lessons from human, mouse and rat genomes}, Genome Research
  14:507--516, 2004.

\bibitem{Bourque} G. Bourque, G. Tesler and P.A. Pevzner, \textit{The
  convergence of cytogenetics and rearrangement-based models for
  ancestral genome reconstruction}, Genome Research 16:311--313, 2006.

\bibitem{Bryant} D. Bryant, \textit{On the uniqueness of the selection
  criterion in neighbor-joining}, Journal of Classification 22:3--15, 2005.
\bibitem{Cai} W. Cai, J. Pei and N.V. Grishin, \textit{Reconstruction of
  ancestral protein sequences and its applications}, BMC Evolutionary
  Biology 4:33, 2004.

\bibitem{BryantMoulton} D. Bryant and V. Moulton, \textit{NeighborNet:
  An agglomerative method for the construction of phylogenetic
  networks}, Evolution 21:255--265, 2004.

\bibitem{Cavender} J. Cavender, \textit{Taxonomy with confidence},
  Mathematical Biosciences 40:271--280, 1978.

\bibitem{Chen} K. Chen and L. Pachter, \textit{Bioinformatics for
  whole-genome shotgun sequencing of microbial communities}, PLoS
  Computational Biology 1:e24, 2005.

\bibitem{Chor} B. Chor and T. Tuller, \textit{Maximum likelihood of
  evolutionary trees is hard}, in Proceedings of RECOMB 2005.

\bibitem{Daskalakis} C. Daskalakis, C. Hill, A. Jaffe, R. Miheascu,
  E. Mossel and S. Rao, \textit{Maximal accurate forests from distance
    matrices}, in Proceedings of RECOMB 2006. 

\bibitem{Dewey} C. Dewey, P. Huggins, K. Woods, B. Sturmfels and
  L. Pachter, \textit{Parametric alignment of Drosophila genomes}, PLoS
  Computational Biology 2:e73, 2006.

\bibitem{DeweyPach} C. Dewey and L. Pachter, \textit{Evolution at the
  nucleotide level: the problem of multiple whole-genome alignment},
  Human Molecular Genetics 15:R51--R56, 2006.

\bibitem{Durrett} R. Durrett and Y. Interian, \textit{Genomic midpoints:
  computationa and evolution implications}, submitted.

\bibitem{Elias} I. Elias and J. Lagergren, \textit{Fast neighbor
  joining}, Proceedings of the International Colloquium on Automata,
  Languages and Programming (ICALP), 2005.

\bibitem{ENCODE} The ENCODE Project Consortium, \textit{The ENCODE
  (ENCyclopedia of DNA Elements) Project}, Science 306:636--640, 2004.

\bibitem{Eriksson} N. Eriksson, K. Ranestad and B. Sturmfels,
  \textit{Phylogenetic algebraic geometry}, in Projective Varieties with
  Unexpected Properties, C. Ciliberto et al. (eds), Water de Gruyter,
  Berlin 2005. 

\bibitem{Fel81} J. Felsenstein, \textit{Evolutionary trees from DNA sequences: a
maximum likelihood approach}, Journal of Molecular Evolution
  17:368--376, 1981.

\bibitem{Felsenstein1981b} J. Felsenstein, \textit{Evolutionary trees from gene
  frequencies and quantitative characters: finding maximum likelihood
  estimates},  Evolution 35:1229--1242, 1981.

\bibitem{Fernandez} D. Fernandez-Baca, B. Venkatachalam, A. Alberto,
  C. Maxime and P. Kunsoo, \textit{Parametric analysis for ungapped
    Markov models of evolution}, in Proceedings of the Conference on
  Combinatorial Pattern Matching, 2005.

\bibitem{GascuelSteel} O. Gascuel and M. Steel, \textit{Neighbor joining
revealed}, Molecular Biology and Evolution 23:1997--2000, 2006.

\bibitem{Gaucher} E.A. Gaucher, J.M. Thomson, M.F. Burgan and
  S.A. Benner, \textit{Inferring the paleoenvironment of ancient
    bacteria on the basis of resurrected proteins}, Nature 425:285--288,
  2003.

\bibitem{Serkan} S. Ho\c{s}ten, A. Khetan and B. Sturmfels, \textit{Solving
  the likelihood equations}, Foundations of Computational Mathematics
  5:389--407, 2005.

\bibitem{Huynh} T.N.D. Huynh, J. Jansson, N.B. Nguyen and W.-K. Sung,
  \textit{Constructing a smallest refining galled phylogenetic network}, in
  Proceedings of RECOMB 2005.
\bibitem{Jojic} V. Jojic, N. Jojic, C. Meek, D. Geiger, A. Siepel,
  D. Haussler and D. Heckerman, \textit{Efficient approximations for
    learning phylogenetic HMM models from data},
  Bioinformatics:20:i161--i168.

\bibitem{Kuang} D. Kuang, Y. Yao, D. MacLean, M. Wang, D.R. Hampson and
  B.S.W. Chang, \textit{Ancestral reconstruction of the ligand binding
    pocket of family C G-protein coupled receptors}, Proceedings of the
  National Academy of Sciences of the USA, 103:14050--14055, 2006.

\bibitem{Kumar} S. Kumar and S.R. Gadagker, \textit{Efficiency of the
  neighbor-joining method in reconstructing evolutionary relationships
  in large phylogenies}, Journal of Molecular Evolution 51:544--553,
  2000.

\bibitem{Lane} D.J. Lane, B. Pace, G.J. Olsen, D.A. Stahl, M.L. Sogin and M.R.
  Pace, \textit{Rapid determination of 16S ribosomal RNA sequences for
    phylogenetics analysis}. Proceedings of the National Academy of
  Sciences, USA 20:6955-6969, 1985.

\bibitem{Liberles} D.A. Liberles (editor), \textit{Ancestral Sequence
  Reconstruction}, Oxford University Press, to appear in 2007.

\bibitem{Ma} J. Ma, L. Zhang, B.B. Suh, B.J. Raney, R.C. Burhans,
  W.J. Kent, M. Blanchette, D. Haussler and W. Miller,
  \textit{Reconstructing contiguous regions of an ancestral genome},
  Genome Research 16:1557--1565, 2006.

\bibitem{McAuliffe} J. McAuliffe, L. Pachter and M.I. Jordan,
  \textit{Multiple sequence functional annotation and the generalized
    hidden Markov phylogeny}, Bioinformatics 20:1850--1860, 2004.

\bibitem{Mihaescu} R.H. Mihaescu, D. Adkins, C. Hill, A. Jaffe and
  S. Rao, \textit{A simple quadratic time algorithm for accurate
    phylogeny reconstruction from logarithmic-sized data}, submitted.

\bibitem{MihaescuPachter} R.H. Mihaescu, D. Levy and L. Pachter,
  \textit{Why neighbor-joining works}, submitted (arXiv cs.DS/0602041).

\bibitem{Needleman} S.B. Needleman and C.D. Wunsch, \textit{A general
  method applicable to the search for similarities in the amino acid
  sequence of two proteins}, Journal of Molecular Biology 48:443--453,
  1970.

\bibitem{Ogurtsov} A.Y. Ogurtsov, S. Sunyaev and A.S. Kondrashov,
  \textit{Indel-based evolutionary distance and mouse-human divergence},
  Genome Research 14:1610--1616, 2004.

\bibitem{Pachterindel} L. Pachter, \textit{The majority of divergence
  between DNA sequences is due to indels}, submitted.

\bibitem{ASCB2005} L. Pachter and B. Sturmfels, \textit{Algebraic
  Statistics for Computational Biology}, Cambridge University Press
  2005.

\bibitem{MathPhyl} L. Pachter and B. Sturmfels, \textit{The mathematics
  of phylogenomics}. SIAM review, in press.


\bibitem{Pauling1} L. Pauling and E. Zuckerkandl, \textit{Chemical
  paleogenetics: molecular restoration studies of extinct forms of
  life}, Acta Chem. Scand. 17:89, 1963.


\bibitem{Pollard} D. Pollard, V.N. Iyer, A.M. Moses and M.B.Eisen,
  \textit{Whole genome phylogeny of the Drosophila melanogaster species
    subgroup: widespread discordance with species tree \& evidence for
    incomplete lineage sorting}, PLoS Genetics, advance access 2006.


\bibitem{Pupko} T. Pupko, A. Doron-Faigenboim, D.A. Liberles and
  G. M. Cannarozzi, \textit{Probabilistic models and their impact on the
    accuracy of reconstructed ancestral sequences}, in Ancestral
  Sequence Reconstruction, D.A. Liberles (editor), Oxford University
  Press, to appear in 2007.

\bibitem{PupkoFast} T. Pupko, I. Pe'er, R. Shamir and D. Graur,
  \textit{A fast algorithm for joint reconstruction of ancestral amino
    acid sequences}, Molecular Biology and Evolution 17:890--896, 2000.

\bibitem{Rocchi} M. Rocchi, N. Archidiacono and R. Stanyon,
  \textit{Ancestral genomes reconstruction: an integrated,
    multi-disciplinary approach is needed}, Genome Research
  16:1441--1444, 2006.

\bibitem{Ross} H.A. Ross, D.C. Nickle, Y. Liu, L. Heath, M.A. Jensen,
  A.G. Rodrigo and J.I. Mullins, \textit{Sources of variation in
    ancestral sequence reconstruction for HIV-1 enevelope genes},
  Evolutionary Bioinformatics 2:18--41, 2006.


\bibitem{NJ} N. Saitou and M. Nei, \textit{The neighbor joining method:
  a new method for reconstructing phylogenetic trees}, Molecular Biology
  and Evolution 4:406--425,1987.

\bibitem{Siepel} A. Siepel and D. Haussler, \textit{Phylogenetic hidden
  Markov models}, in Statistical Methods in Molecular Evolution
  (R. Nielsen editor), Springer, NY, p 325--351, 2005.

\bibitem{Steel} C. Semple and M. Steel, \textit{Phylogenetics}, Oxford
  University Press, 2003.

\bibitem{Snir} S. Snir and L. Pachter, \textit{Phylogenetic profiling of
  insertions and deletions}, in Proceedings of RECOMB 2006.

\bibitem{BerndClay} B. Sturmfels, \textit{Can biology lead to new
  theorems?}, Annual report of the Clay Mathematics Institute, 2005.

\bibitem{Thomas} J.W. Thomas et al., \textit{Comparative analysis of
  multi-species sequences from targeted genomic regions}, Nature
  424:788--783, 2003.

\bibitem{Thorntonrev} J.W. Thornton, \textit{Resurrecting ancient genes:
  experimental analysis of extinct molecular}, Nature Reviews Genetics
  5:366--375, 2004.

\bibitem{Thornton} J.W. Thornton, E. Need and D. Crews,
  \textit{Resurrecting the ancestral steroid receptor: ancient origin of
    estrogen signaling}, Science 301:1714--1717, 2003.

\bibitem{Wang} L.-S. Wang and T. Warnow, \textit{Reconstructing
  chromosomal evolution}, SIAM Journal on Computing, in press.

\bibitem{Lusheng} L. Wang and D. Gusfield, \textit{Improved
  approximation algorithms for tree alignment}, Journal of Algorithms
  25:255--273, 1997.

\bibitem{Williams} P.D. Williams, D.D. Pollock, B.P. Blackburne and
  R.A. Goldstein, \textit{Assessing the accuracy of ancestral protein
    reconstruction methods}, PLoS Computational Biology, advance access
  (2006).

\bibitem{Woese} Woese, C.R. and GE Fox, \textit{Phylogenetic structure
  of the prokaryotic domain: the primary
  kingdoms}. Proceedings of the National Academy of Sciences, USA 74:5088--5090.

\bibitem{Yandell} M. Yandell, C.J. Mungall, C. Smith, S. Prochnik,
  J. Kamniker, G. Hartzell, S. Lewis and G.M. Rubin, \textit{Large-scale
    trends in the evolution of gene structures within 11 animal
    genomes}, PLoS Computational Biology 2:e15, 2006.

\bibitem{Yap} V.B. Yap and L. Pachter, \textit{Identification of
  evolutionary hotspots in the rodent genomes}, Genome Research
  14:574--579, 2004.

\bibitem{Pauling2} E. Zuckerkandl and L. Pauling, \textit{
Molecular disease, evolution, and genetic heterogeneity}, in M. Kahsa
  and B. Pullman eds. Horizons in biochemistry. Academic Press, New
  York. 1962.

\bibitem{Pauling3} E. Zuckerkandl and L. Pauling, \textit{
Molecules as documents of evolutionary history}. Journal Theorertical
  Biology 8:357--366, 1965.



\end{thebibliography}

\end{document}